\documentclass[]{spie}  %>>> use for US letter paper
\usepackage{amsmath,amsfonts,amssymb}
\usepackage{graphicx}
\usepackage[colorlinks=true, allcolors=blue]{hyperref}

\title{General-Purpose Software for Managing Astronomical Observing Programs in
the LSST Era}

\author[a]{R.A.~Street, M.~Bowman, E.S.~Saunders, T.~Boroson}
\affil[a]{LCOGT, 6740 Cortona Drive, Suite 102, Goleta, CA 93117, USA}
\authorinfo{Further author information: (Send correspondence to R.A.~Street)\\R.A.~Street: E-mail: rstreet@lco.global, Telephone: 1 805 880 1631}

\begin{document} 
\maketitle

\begin{abstract}
Modern astronomical surveys such as the Large Synoptic Sky Survey (LSST) promise an unprecedented wealth of discoveries, delivered in the form of ~10 million alerts of time-variable events per night.  Astronomers are faced with the daunting challenge of identifying the most scientifically important events from this flood of data in order to conduct effective and timely follow-up observations.

Several ongoing observing programs have proven databases to be extremely valuable in conducting efficient follow-up, particularly when combined with tools to select targets, submit observation requests directly to ground- and space-based facilities (manual, remotely-operated and robotic), handle the resulting data, interface with analysis software and share information with collaborators.  We draw on experience from a number of follow-up programs running at LCOGT, all of which have independently developed systems to provide these capabilities, including the Microlensing Key Project (RoboNet, PI: Tsapras, co-I Street), the Global Supernova Project (SNEx, PI: Howell) and the Near-Earth Object Project (NEOExchange, PI: Lister).  We refer to these systems in general as Target and Observation Managers (TOMs).  

Future projects, facing a much greater and rapidly-growing list of potential targets, will find such tools to be indispensable, but the systems developed to date are highly specialized to the projects they serve and are not designed to scale to the LSST alert rate. 

We present a project to develop a general-purpose software toolkit that will enable astronomers to easily build TOM systems that they can customize to suit their needs, while a professionally-developed codebase will ensure that the systems are capable of scaling to future programs.
\end{abstract}


\section{INTRODUCTION}
\label{sec:intro}  % \label{} allows reference to this section

Developments in telescope, instrument and computing technologies have enabled tremendous improvements in the capabilities of astronomical surveys.  It is now possible to survey the night's sky to higher spatial resolutions, fainter brightness limits and larger sky area than ever before, and to repeat the observations night after night.  This has opened up the hitherto under-explored time-domain, and vast new catalogs of phenomena that vary in brightness and/or position are anticipated from near-future survey facilities including the Zwicky Transient Facility (ZTF \cite{ZTF} \footnote{http://www.ztf.caltech.edu/}) and the Large Synoptic Survey Telescope (LSST \footnote{http://lsst.org}).  Combined with rapid data reduction and analysis, these facilities will also issue a stream of alerts for new and changing targets.  

These new data streams are expected to deliver a wealth of new discoveries \cite{LSSTSciBookv2} from Trans-Neptunian Object binaries to Active Galactic Nuclei.  Yet for many phenomena, the survey data alone will be insufficient to properly classify and characterize the nature of the target.  Astronomers will  have to orchestrate a program of follow-up observations in order to fullfill their science goals.  In many cases, particularly for transient phenomena such as supernovae, tidal disruption events and moving objects, follow-up observations will have to be made in rapid reaction to the initial discovery alert, since the source of the alert which triggered it will become unavailable after some period of time, leaving the discovery incompletely understood.  

As long as the rate of new target discovery within a given field of astronomy remained relatively low, the selection of targets for follow-up observations could be done by an astronomer or a team, who could also then make the observations and analyse the resulting data in a timely manner.  But in many fields of astronomy, target discovery rates have already exceeded the point at which this approach becomes unsustainable, leading to great inefficiency, and potentially important discoveries remaining poorly constrained.  Within those fields, several teams have turned to software to automate as much of the follow-up program as possible, demonstrating the effectiveness of such tools, which are referred to by the generic name of Target and Observation Manager systems, or TOMs. With the advent of near-future surveys, these tools are likely to be needed by all areas of astronomy: ZTF is expected to produce up to $\sim$1\,million alerts per night, and LSST may produce $\sim$10\,million.  

However, the systems that currently exist are heavily specialised to their original science goals, and require a high degree of expertise in database management and software development that is normally outside the training of astronomers.  Furthermore they are not expected to scale to the large target volume expected from near-future surveys.  Here we describe a project to develop a general-purpose software `toolkit' which will enable astronomers to easily build their own TOM system and customize it to the needs of their science. 

In Section~\ref{sec:followup} we describe the common elements of follow-up observing programs, in the context of the inter-related facilities that they depend on,  as a prelude to outlining the functional requirements of a TOM system, and briefly summarize the key features of some existing TOMs in Section~\ref{sec:currenttoms}.  We then describe the goals and design of the TOM Toolkit in Section~\ref{sec:toolkit} and plans for its implementation and community engagement in Section~\ref{sec:workplan}.  

\section{CONDUCTING FOLLOW-UP OBSERVING PROGRAMS}
\label{sec:followup}

\subsection{The Role of a TOM in the Time-Domain Astronomy Network}

\begin{figure} [ht]
   \begin{center}
   \begin{tabular}{c} 
   \includegraphics[height=14cm]{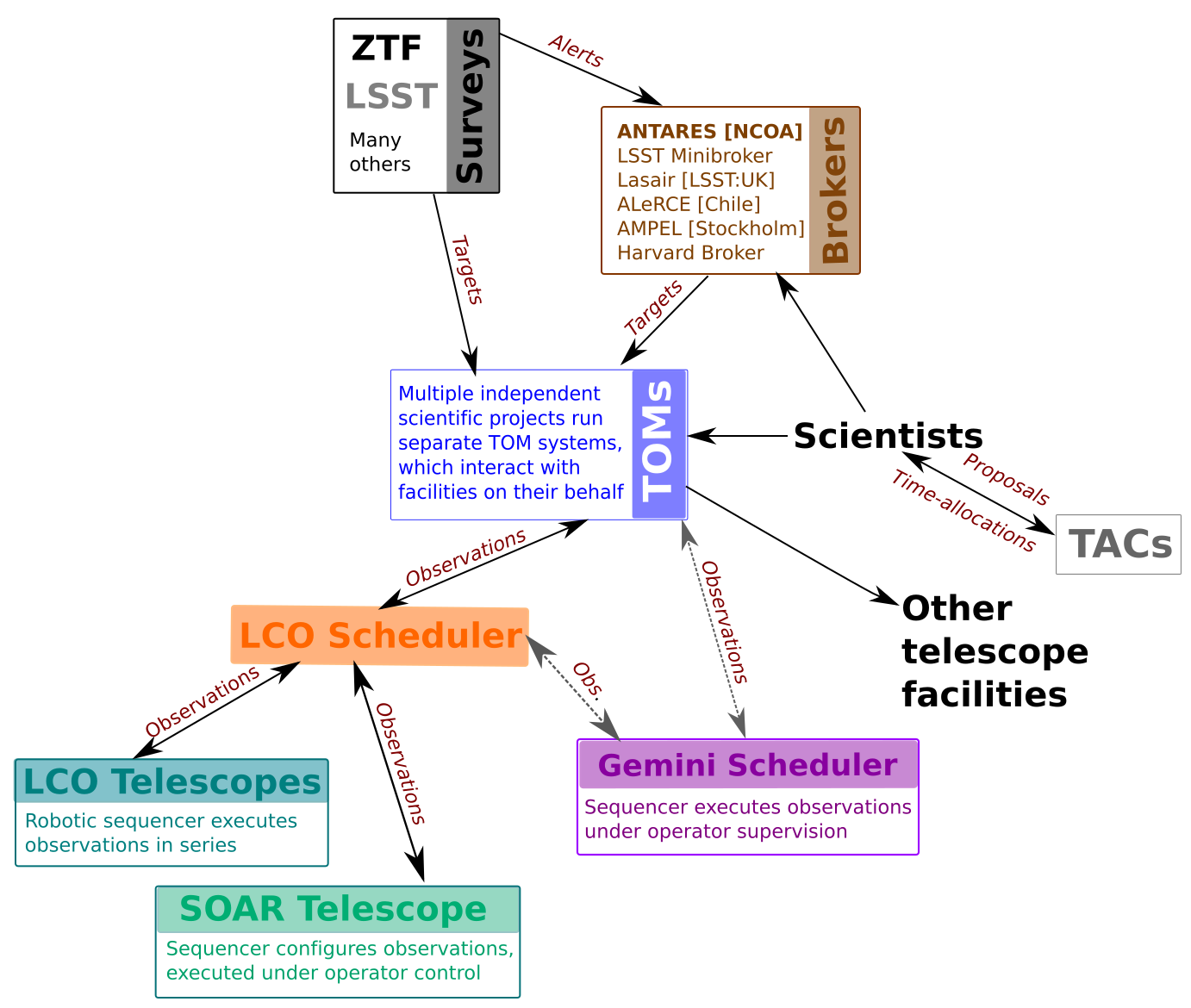}
   \end{tabular}
   \end{center}
   \caption[example] 
   { \label{fig:tda-network} 
The `ecosystem' for time-domain astronomy - relationships between the key facilities necessary to conduct large-scale follow-up observation programs in response to survey alerts.   Solid lines indicate existing links or those currently under active development; dashed lines indicate areas of likely future development.}
   \end{figure}

Time-domain astronomy relies upon a chain of inter-related facilities and services; the relationships between these systems is shown in Figure~\ref{fig:tda-network}.  In general each link in the chain is developed and operated by an independent group, but depends on well-defined interfaces with adjacent services working smoothly.  This is sometimes referred to as the time-domain network, or ecosystem.  

Alerts of new or changing targets are provided by a number of services and widely disseminated.  Broker services cross-match these alerts with existing catalog data, for example photometry at other wavelengths not covered by the originating survey, spectra, or information on the object's past behavior.  Many brokers then use the collated data to determine a more informed classification for each alert, to enable scientists to select targets of particular importance to their programs from a much wider variety of time-variable phenomena.  Brokers commonly provide online interfaces, designed for human and/or software interactions, from which their data on each object can be browsed and extracted.  Some offer more sophisticated query tools to enable further sub-selection of targets.  

TOM systems are designed to harvest information on targets of interest, often from a range of sources including pre-existing catalogs as well as interfacing directly with survey alert streams and broker services.  The result is a combined database of targets of specific interest to the science goals of the project which runs the TOM system.  The scientist must then choose which targets require additional observations and from what instrument and telescope; this can be done manually or by an algorithm which can submit and retrieve data to and from the TOM.  Sometimes the observations are carried out independently of the TOM software, but several TOM systems have been designed which can submit observation requests directly, and even automatically, to compatible telescope facilities.  For transient targets especially, it is often necessary to rapidly access and reduce the data products from the observations, because the outcome can determine the relative priority of the target and on what, if any, observations happen next.  For this reason, TOM systems frequently interface with data archives and reduction software in order to facilitate the rapid data analysis.  

\subsection{Requirements of Time-Domain Follow-up Programs}
In order to design software for a TOM system, we first reviewed the practical requirements of a range of scientific programs that require follow-up.  In particular we considered the following use-cases:

\begin{enumerate}
\item {\bf Transient targets (e.g. supernovae, tidal disruption events, microlensing events, flares)}\\
A diverse range of astrophysical targets exhibit unique, unpredictable non-repeating phenomena, including explosive transients such as supernovae, eruptive variability such as stellar flares and microlensing.  Due to their intrinsic nature, transient targets are usually identified by surveys that issue alerts in real-time and a premium is frequently placed on obtaining observations very rapidly.  However transient targets normally also require monitoring over a range of timescales from days to years at a cadence that may vary depending on the evolving behavior of each object.  This diverse category of phenomena requires a similarly broad range of telescope facilities including multi-band photometry at all wavelengths and spectroscopy at all resolutions, so observations must be coordinated across a range of instruments and telescopes including space-based facilities and gravitational wave detectors.  \\

\item {\bf Periodic and quasi-periodic objects (e.g. transiting exoplanets, pulsating stars, eclipsing systems)}\\
A number of astrophysical phenomena occur repeatedly, usually as a result of rotation, orbital motion or pulsations.  Though there may be multiple opportunities for observations, they must often be timed to occur within strict windows coinciding with specific phases such as eclipses and are often required to be intensive within those windows.  For example, a single target requires an extended sequence of back-to-back exposures for the duration of the window.  It can also be scientifically valuable to conduct simultaneous observations with multiple instruments in order to better understand the object’s behavior.   Given the restricted time windows during which targets can be observed, careful planning is required to optimize those opportunities.  A number of planning tools are commonly used for this purpose, e.g. plots of the altitude of a target above the horizon of a given observing facility, and the corresponding airmass, within a specified date range, plots of the angular separation between the target and the Moon, also representing the Moon phase,  as a function of time and a calendar display of the opportunities to observe primary and secondary eclipses of a given target from a given set of observatories.  
The most common observation types are multi-band optical and NIR photometry (normally ~450-2400nm) as well as medium-high resolution spectroscopy (R$\sim$40,000+, from both long-slit and echelle instruments) and adaptive-optics imaging.  Targets are usually derived from large survey catalogs, and data are typically gathered and subsequently reduced over the course of weeks to months.  

\item {\bf Moving Objects (e.g. Near-Earth Asteroids (NEOs), dwarf and major planets)}\\
Objects within our Solar System and stars and brown dwarfs within a few parsecs of the Sun show appreciable change in their position over the timescales of a follow-up program (up to years).  They require predominantly optical and NIR imaging in order to measure astrometry and photometry, though long-slit optical spectroscopy is also required.  As their positions must be computed for the exact time of the observation (and may require non-sidereal tracking), observers need to specify the object in terms of its ephemeris parameters (Solar System targets) or coordinate reference and proper motion (nearby Galactic targets).  While some objects are likely to return, follow-up astrometry must be obtained within ~2 days of discovery in order to determine the object’s solar orbit precisely enough to track it in future.  Targets can be ‘lost’ so rapidly that data must be reduced and analyzed immediately, and all measurements promptly shared publicly.  

\item {\bf Large-sample surveys (e.g. Galaxy classification spectra, brown dwarf taxonomy survey, mission input catalog preparation)}\\
As astronomical surveys expand in capabilities, our catalogs of objects grow apace.  Some projects require just a single observation per object, but depend on compiling consistent observations of a large sample of objects to achieve their science goal.  Many of the projects in this category have no time constraint on when the observations are acquired, and less urgency to receive the data than the other categories, but nevertheless depend on TOM systems simply to keep track of the large volume of targets, observations and data. 
\end{enumerate}

Drawing on these use-cases, it is evident that a TOM system must interface with a variety of surveys, catalogs and alert streams, and provide the means for users to select from those sources candidate targets of interest to the user's specific science goals.  While support for every possible data source would be a daunting task, there are a finite number of such sources that are widely used within astronomy, so  providing tools to interface with those sources enables a wide range of science.  These include, for example, the ANTARES broker\footnote{https://www.noao.edu/ANTARES/}\cite{Saha2014, Saha2016} (which is expected to handle both the ZTF and LSST alert streams), the Minor Planet Center\cite{MPC} (a well-established broker for Solar System objects) and the TESS Input Catalog\cite{TIC} (a catalog of stars to be studied by the Transiting Exoplant Survey Satellite).  

In all of these use-cases, the number of available targets already exceeds the number for which adequate follow-up observations could be obtained with the telescope facilities available.  This situation can only get worse with more survey facilities coming online, so astronomers must prioritise their candidates and select only the most scientifically valuable.  Every science project will have distinctive scientific goals, so the criteria for this selection will vary in every case, so astronomers must be able to apply their own selection through the TOM system, by manual interaction with the database and/or by interfacing their own selection software with the TOM.  

Some of the use cases require that observations be taken of a given object over an extended period, up to years in some cases.  Yet more targets are discovered all the time.  This leads to a continuous re-evaluation of relative target priorities, given finite telescope facilities on which to follow them up.  It also implies the need to keep track of observations of existing targets while simultaneously assessing the demands for new  ones.  This process must often take into account limited availability windows for observing resources, for example if a particular instrument may only be mounted for specific periods.  

Follow-up for a single object may encompass a diverse range of telescope facilities, each of which has their own Time Allocation Committee (TAC).  We proceed on the assumption that the project running a TOM system will apply to the TACs of all relevant facilities in advance, and assume that time has been granted.  This mode of operation may change in the future (a possibility we discuss in the conclusions) but it is likely that some authentication will be necessary in order for observations to be requested from a given resource.  Facilities are becoming increasingly roboticized and many are already able to receive robotically-submitted observation requests via Application Programming Interfaces (APIs) \cite{Saunders2018}.  This capability is particularly valuable for transient targets, so it leads to the requirement that the TOM should be configurable with the authentication details of time awarded on relevant facilities.  Wherever possible, the TOM should also enable users to design and submit observing requests directly to those facilities.  Again, this requires the TOM to support a range of APIs, but it is a finite range for the major common-user telescopes.  

Similarly, when astronomers attempt to access the data resulting from their observations, they will do so through a finite set of archive facilities, which are also increasingly accessible through APIs.  By providing TOM tools to enable users to access their data in these archives, it becomes possible for users to build their own automated data handling pipelines.  This is an essential labor-saving component for transient programs that need to conduct real-time analysis in order to decide future priorities and observations.  
It is also a requirement that the users should be able to plug their own software into the data reduction pipeline, since methods of analysis improve all the time, and each user is likely to need to use bespoke algorithms.  

There is a tendency to think of surveys as ``large datasets'' and to somewhat dismiss follow-up observations as a comparatively small effort.  While this is generally true, the scale of follow-up (and hence the resources required) should not be underestimated.  To give an illustrative example, the first LCO Microlensing Key Project ran from 2014-2017, obtaining follow-up optical imaging of microlensing events between February -- October each year (due to target visibilty).  During that period, the project made 2,631 discrete observation requests, leading to $\sim$15,000 images, at $\sim$90\,MB each.  Thus the raw follow-up data alone amounted to $\sim$12\,TB while the ``reduced'' data products took an additional $\sim$30\,TB.  This represents a significant effort in data curation and simply keeping track of the status of the reduction of each frame and dataset.  
 
Finally, a great deal of science is now carried out by large collaborations, often including members distributed across several countries and timezones.  This can be extremely useful in carrying out observing programs that must respond 24/7, but leads to some logistical issues.  All team members must be able to access the information they need to perform their aspect of the program, and the TOM infrastructure should facilitate this.  

\subsection{Data Sharing}
Sharing data products before publication is a thorny issue in astronomy.  In some cases, it is necessary to achieve the science goals, particularly for transient targets where no single group has the resources to obtain the full set of data required.  Yet there are often concerns that sharing data can lead to another team `scooping' future publications which can have real impacts on careers and even funding streams.  As a result, different fields in astronomy have developed quite different expectations regarding what data can be shared and at what stage. 

For example, in the NEO community all candidates are shared publicly via the broker services provided by the Minor Planet Center, which also serves to combine data products submitted very soon after they are taken by the community in order to update the orbital parameters for each object as quickly as possible.  In contrast, exoplanet searches have routinely restricted access to their data products and even the target list of objects they intend to study.  Interestingly, this culture gradually changed when presented with a large catalog of targets from the {\em Kepler} Mission, when it was clear all available resources would still take years to follow-up all possible targets.  It has now become the norm to share data products via systems such as NASA's Exoplanet Follow-Up Observing Program (ExoFOP\cite{ExoFOP}).  

For the TOM Toolkit to be broadly adopted by the community, it must have the functionality to share data {\it if the user wishes to,} and provide sufficiently fine-grained controls that teams can implement the data sharing model they choose to adopt. 

\section{EXAMPLES OF CURRENT TOM SYSTEMS}
\label{sec:currenttoms}

The need for TOM systems has been recognized for some time \cite{WhiteAllan2008}, and examples can be found dating back at least a decade as numerous projects across astronomy have developed software customized to their goals.  A more extensive list was presented by \citenum{Arcavi2017}; here we include a few representative examples to highlight the range of capabilities and to provide context for the folllowing sections.  

\begin{itemize}
\item {\bf Palomar Transient Factory (PTF) Follow-up Marshall}
A database developed to hold and present all available data on targets identified in the optical imaging and photometry from the PTF survey\cite{Rau2009PTF, Law2009PTF}.  Target classifications were provided to the database by the RealBogus machine learning algorithm\cite{Miller2015}, to help project members choose suitable targets for further study.  The Marshall provided an online portal interface for geographically-distributed members, as well as the option for users to upload additional data products from follow-up observations and a forum where members could post comments regarding a given target.  

\item {\bf NASA Exoplanet Follow-Up Portal (ExoFOP\cite{ExoFOP})}
This database was originally developed by IPAC to help the exoplanet community to coordinate follow-up observations of {\em Kepler} transiting planet candidates, and has since been expanded to support microlensing planet searches.  It provides searchable lists of all available candidates, crossed-matched against catalog data, with displays summarizing the information available for each object.  These include all available data products, and users are encouraged to contribute observation products to the publicly available data.  The ExoFOP provides a number of tools to help with common observation planning and data analysis tasks including calculating what transit events can be observed from a given telescope on a given date, and lightcurve periodogram analysis.  

\item {\bf Supernova Exchange (SNEx\cite{SNEx})}
SNEx collates information on supernovae (and other events e.g. tidal disruption events, kilonovae, supernova impostors) discovered by a range of surveys, presenting the user with an interactive summary display of all current information, including plots of follow-up data products that are accessible by the user and posted user comments.  SNEx offers the option for users to easily compose and directly submit requests for optical imaging and spectroscopy from the LCO Network.  As the supernova field is comprised of a large number of independent teams, some of whom opt to share some of their data, SNEx enables many different teams to use the same TOM system by offering team leaders the option to choose whether they share any given observation and data product, and with whom. 

\item {\bf RoboNet Microlensing System \cite{RoboNet,Tsapras2009RBN}}
This TOM automatically harvests target information both from alert-issuing microlensing surveys and from ARTEMiS, a broker service designed to collate available data on microlensing events and detect anomalies in the lightcurves.  Though the RoboNet system provides an online interactive portal, it was designed to operate without human supervision, to ensure as rapid response as possible.  The TArget Prioritization (TAP \cite{Hundertmark2018}) software automatically selects events for observations that are robotically submitted directly to the LCO Network.  It has also previously conducted robotic observations on the Liverpool Telescope.  The system includes support for a fully robotic Difference Image Analysis (DIA) data pipeline to harvest and analyse the observations, and to use the results to guide future target selection.  The data pipeline is currently being re-written in Python and C++ to avoid the financial burden of licensing the original IDL software (DanDIA \cite{Bramich2008, DanDIA, Bramich2013}), to optimize specific time-consuming algorithms for speed and to parallelize wherever possible.  The system provides a number of online displays for users to ensure the software was running normally, inspect summaries of the available data on a given object, as well as interactive tools for users to control the system via the website.  

\item {\bf NEOExchange \cite{NEOExchange}}
NEOExchange receives alerts of NEO candidates from a number of surveys via the Minor Planet Center as well as from other priority targets from radar assets and NASA. A customized algorithm draws from the NEOExchange database and optimizes targets selected for follow-up. This factors in time since last observation and length of observed arc, potential risk to the Earth as well as observability from the telescopes in the LCO Network. It provides interactive tools to enable users to submit spectroscopic and imaging observation requests to the LCO Network, as well as displays of the available information and a list of current targets.  Data taken for the submitted requests is recorded in NEOExchange along with some quality control information from the frames. The catalog of sources detected in the frames along with any moving object candidates are also ingested. This allows the examination and confirmation of NEO candidates in the browser and also allows for automated light curve construction for moving targets.
\end{itemize}

\section{THE TOM TOOLKIT}
\label{sec:toolkit}

\begin{figure} [ht]
   \begin{center}
   \begin{tabular}{c} 
   \includegraphics[height=9cm]{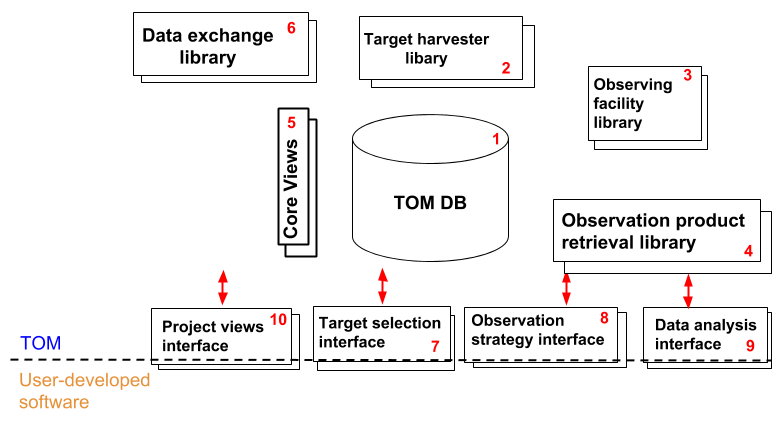}
   \end{tabular}
   \end{center}
   \caption[example] 
   { \label{fig:tom-schematic} 
A schematic overview of the TOM Toolkit components and how they are envisioned to interface with user-developed software.}
   \end{figure}

The goals of the TOM Toolkit package are to develop a general-purpose software toolkit that will enable astronomers to easily build TOM systems that they can customize to suit their needs, using a professionally-developed codebase that will ensure that the systems are capable of scaling to future programs.  Since almost all science programs are likely to develop their own software and approach for target selection, data analysis and other aspects of their projects, the Toolkit is designed to provide the fundamental components of a TOM in such a way that they can be easily built into the system the user requires, rather than attempt to deliver a single, `one-size-fits-all' package.   Figure~\ref{fig:tom-schematic} gives an overview of the elements of the Toolkit and illustrates how they are expected to interface with user-developed software.  The Toolkit will consist of the following libraries and interfaces.  

\begin{enumerate}
\item {\bf The TOM Database (TOM-DB)}
The TOM database needs to be flexible enough to store two distinct types of data.  The Core Target Data will include the essential parameters needed to describe and identify astronomical targets including multiple identifiers, position and timing together with commonly-used science-specific parameters.  The TOM-DB will also store Science Target Documents in a data format specified by the user, enabling them to include any information relevant to their science.  

\item{\bf Target Harvester Library (THL)}
This library will include functions that harvest target information from commonly-used sources and store it in the TOM DB.  Examples include the Transient Name Server, Astronomer’s Telegrams, the Minor Planet Center, TESS, ZTF, the ARTEMiS system, and these will be used to define an API to make the adoption of future alert services more standardized.  The THL will also provide query tools for common-user services and archives.  

\item {\bf Observing Facility Interface (OFI)}
The OFI will provide tools to allow users to specify the parameters of observation requests for supported instruments and facilities, and to submit those requests to the facilities. It will also provide functions to receive information from the facilities regarding the current status of the observations and the facilities themselves (e.g. has the observation been made?  Is the telescope offline due to poor weather?).  The reference implementation of the OFI will be to support observation requests to the LCO Network and the SOAR Telescope, and it is hoped that the resulting API definition will be easily extensible to other facilities \cite{Saunders2018}.  

\item {\bf Observation Product Retrieval Library (OPRL)}
This library will provide functions to programmatically access online archives of data from observing facilities, and will enable users to automatically download their data, and store descriptive parameters in the TOM-DB.  

\item {\bf Core Views Library (CVL)}
Many existing TOM systems have several display tools in common, for example summary displays of essential information on a given target, plotting tools for common data products such as lightcurves and spectra, lists of targets, observation summaries, common query tools such as search around a given sky location and observation-specification forms.  The Toolkit CVL will provide these ready-built to enable users to easily design informative displays of their project's data.  It will also provide tools for user account and data access management.  

\item {\bf Data Exchange Library}
This library will enable users to (optionally) share information on targets, lists of selected targets and data products with other TOM systems and external services.  

\item {\bf Target Selection Interface (TSI)}
It is expected that each project will implement their own preferred method of selecting targets, which may depend on algorithms or human assessment or a combination of both.  The TSI will provide programmatic interfaces to allow users to easily read and query relevant target information from the TOM-DB (both Core and Science Target Data), to write to it parameters computed by the user's software and to add and remove targets from target lists within the TOM-DB.  

\item {\bf Observation Strategy Interface (OSI)}
Similarly, users are expected to define their own observations.  The OSI will provide observation template forms for supported facilities, and tools to enable users to complete and submit them programmatically.  

\item {\bf Data Analysis Interface (DAI)}
Once new observations are obtained, the Toolkit assumes that data will be reduced and analysed by the observatory and the user, using software external to the TOM itself, since there are many excellent packages for this purpose already.  Instead, the Toolkit provides the means for users to input to the TOM-DB data and parameters resulting from that analysis.  The DAI will also work in conjuction with the OPRL to trigger the download of data for analysis.  

\item {\bf Project Views Interface (PVI)}
While the CVL will provide many of the essential elements of an online TOM interface, it is likely that projects will want to develop specialized interfaces and tools for their science.   The PVI will allow them to do so, providing query and update tools for the Core Data and Science Target Documents.   Users will be able to combine these with Core Views to build powerful and highly flexible customized online interfaces.  
\end{enumerate}

Once completed, the TOM Toolkit package will include full documentation of all functions, as well as a simple example of a TOM system, which users can adopt or adapt for their own purposes.  

\section{IMPLEMENTATION AND COMMUNITY ENGAGEMENT}
\label{sec:workplan}

\begin{figure} [ht]
   \begin{center}
   \begin{tabular}{c} 
   \includegraphics[width=17cm]{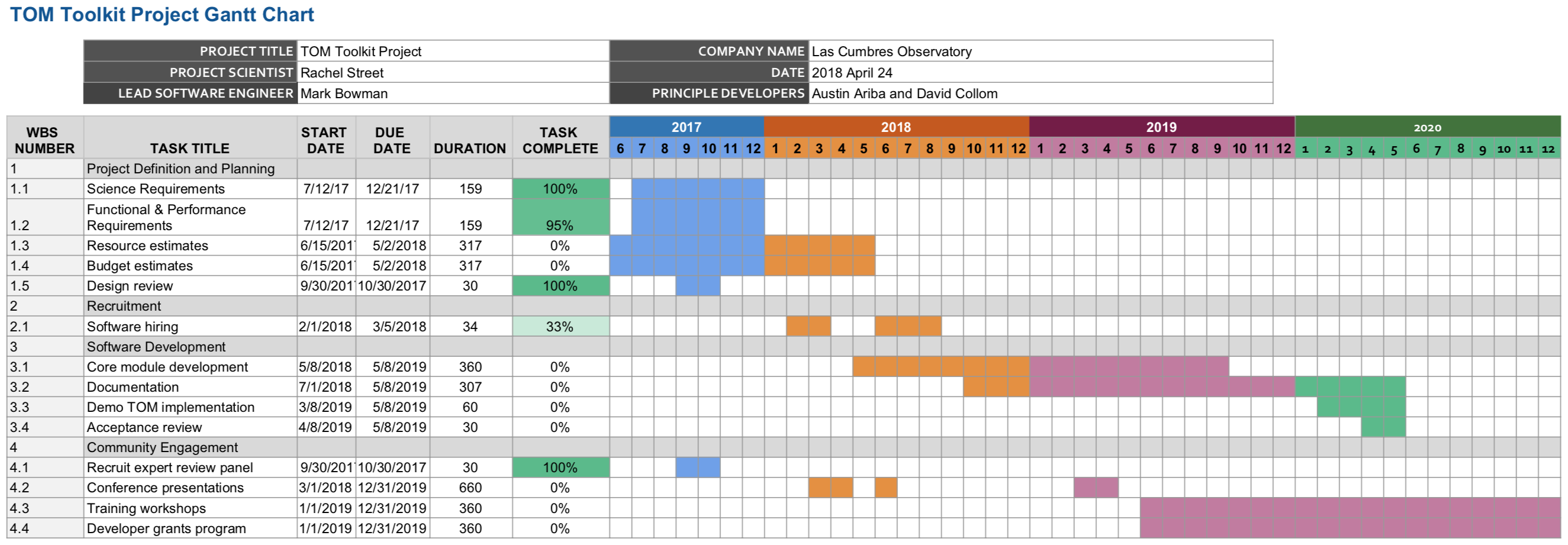}
   \end{tabular}
   \end{center}
   \caption[example] 
   { \label{fig:gantt-chart} 
Gantt chart for the development of the TOM Toolkit.}
   \end{figure}

The Scientific and Functional and Performance Requirements Documents for the TOM Toolkit project are complete, and were reviewed by an invited panel of community experts in October 2017.  As of May 2018, the project has begun hiring for the software engineers who will develop the core libraries and interfaces of the minimum viable product by mid-2019, and the complete product by mid-2020.  Figure~\ref{fig:gantt-chart} shows the expected development timeline for the project.  

For the Toolkit to be successful, it must of course be adopted by the user community, other observatories and institutions.  To increase future user awareness, and to engage them in development, we plan to hold introduction and tutorial sessions at major astronomy conferences and to host a workshop for users to work closely with developers.  

\section{CONCLUSIONS}
This project will develop a software toolkit that will enable astronomers to build systems capable of handling the targets, observations and data products of the large-scale follow-up programs that will be necessary in order to fully realise the scientific potential of near-future surveys such as ZTF and LSST.  The package will be designed to interface seamlessly with other software tools currently under development, including the ANTARES Broker service, and the telescope facilities joining the Time-Domain Astronomy Network, particularly the LCO Network, SOAR and Gemini.  

Networking telescope facilities in this way holds great potential for optimizing and highly efficient follow-up of a wide range of astronomical phenomena, but it also raises the option to shift some long-established paradigms.  For example, should telescope time allocations continue to be made by multiple independent committees, requiring scientists to submit several coordinated proposals, or would a joint committee be beneficial?   With the SOAR and Gemini telescopes acting as pathfinders, APIs for observation requests and observatory status information will be developed as a necessary part of coordinating those facilities with the LCO Network.  Once completed, these APIs will be published, and discussions invited with other telescope operators to investigate expanding it to support other facilities and instruments.  

\section{ACKNOWLEDGEMENTS}
This project receives partial support from a generous grant from the Zegar Family Foundation.

% References

%\bibliographystyle{spiebib} % makes bibtex use spiebib.bst


\begin{thebibliography}{} 
\bibitem{ZTF}{{Smith}, R.~M. and {Dekany}, R.~G. and {Bebek}, C. and {Bellm}, E. and 
	{Bui}, K. and {Cromer}, J. and {Gardner}, P. and {Hoff}, M. and 
	{Kaye}, S. and {Kulkarni}, S. and {Lambert}, A. and {Levi}, M. and 
	{Reiley}, D.}, 2014, {Proc. SPIE}, 9147, 914779.
\bibitem{LSSTSciBookv2}{{LSST Science Collaboration} and {Abell}, P.~A. and {Allison}, J. and 
	{Anderson}, S.~F. and {Andrew}, J.~R. and {Angel}, J.~R.~P. and 
	{Armus}, L. and {Arnett}, D. and {Asztalos}, S.~J. and {Axelrod}, T.~S. and et al.}, 2009, arXiv: {0912.0201}.
\bibitem{Saha2014} {{Saha}, A. and {Matheson}, T. and {Snodgrass}, R. and {Kececioglu}, J. and 
	{Narayan}, G. and {Seaman}, R. and {Jenness}, T. and {Axelrod}, T.},  2014, {Proc. SPIE}, 9149, 914908. 
\bibitem{Saha2016} {{Saha}, A. and {Wang}, Z. and {Matheson}, T. and {Narayan}, G. and 
	{Snodgrass}, R. and {Kececioglu}, J. and {Scheidegger}, C. and 
	{Axelrod}, T. and {Jenness}, T. and {Ridgway}, S. and {Seaman}, R. and 
	{Taylor}, C. and {Toeniskoetter}, J. and {Welch}, E. and {Yang}, S. and 
	{Zaidi}, T.},  2016, {Proc. SPIE},  9910, 99100F.  
\bibitem{MPC}{{Holman}, M. and {Williams}, G. and {Rudenko}, M. and {Keys}, S. and {Payne}, M.},{https://www.minorplanetcenter.net/}.
\bibitem{TIC} {{Stassun}, K.~G. and {Oelkers}, R.~J. and {Pepper}, J. and {Paegert}, M. and 
	{De Lee}, N. and {Torres}, G. and {Latham}, D. and {Muirhead}, P. and 
	{Dressing}, C. and {Rojas-Ayala}, B. and {Mann}, A. and {Fleming}, S. and 
	{Levine}, A. and {Silvotti}, R. and {Plavchan}, P. and {the TESS Target Selection Working Group}}, 2017, arXiv: 1706.00495.
\bibitem{Saunders2018} {{Saunders}, E.S. and {et al.}}, 2018, {Proc. SPIE}, 10704-37. 
\bibitem{ExoFOP}  {{NASA Exoplanet Follow-up Observing Program}}, {https://exofop.ipac.caltech.edu/}.
\bibitem{WhiteAllan2008} {{White}, R.~R. and {Allan}, A.}, 2008, AN, 329, 232.
\bibitem{Arcavi2017} {{Arcavi}, I.},l{https://www.noao.edu/meetings/lsst-tds/presentations/Arcavi\_TOMexamples.pdf}. 
\bibitem{Rau2009PTF}  {{Rau}, A. and {Kulkarni}, S.~R. and {Law}, N.~M. and {Bloom}, J.~S. and 
	{Ciardi}, D. and {Djorgovski}, G.~S. and {Fox}, D.~B. and {Gal-Yam}, A. and 
	{Grillmair}, C.~C. and {Kasliwal}, M.~M. and {Nugent}, P.~E. and 
	{Ofek}, E.~O. and {Quimby}, R.~M. and {Reach}, W.~T. and {Shara}, M. and 
	{Bildsten}, L. and {Cenko}, S.~B. and {Drake}, A.~J. and {Filippenko}, A.~V. and 
	{Helfand}, D.~J. and {Helou}, G. and {Howell}, D.~A. and {Poznanski}, D. and 
	{Sullivan}, M.}, 2009, PASP, 121, 1334.
\bibitem{Law2009PTF} {{Law}, N.~M. and {Kulkarni}, S.~R. and {Dekany}, R.~G. and {Ofek}, E.~O. and 
	{Quimby}, R.~M. and {Nugent}, P.~E. and {Surace}, J. and {Grillmair}, C.~C. and 
	{Bloom}, J.~S. and {Kasliwal}, M.~M. and {Bildsten}, L. and 
	{Brown}, T. and {Cenko}, S.~B. and {Ciardi}, D. and {Croner}, E. and 
	{Djorgovski}, S.~G. and {van Eyken}, J. and {Filippenko}, A.~V. and 
	{Fox}, D.~B. and {Gal-Yam}, A. and {Hale}, D. and {Hamam}, N. and 
	{Helou}, G. and {Henning}, J. and {Howell}, D.~A. and {Jacobsen}, J. and 
	{Laher}, R. and {Mattingly}, S. and {McKenna}, D. and {Pickles}, A. and 
	{Poznanski}, D. and {Rahmer}, G. and {Rau}, A. and {Rosing}, W. and 
	{Shara}, M. and {Smith}, R. and {Starr}, D. and {Sullivan}, M. and 
	{Velur}, V. and {Walters}, R. and {Zolkower}, J.}, 2009, PASP, 121, 1395.
\bibitem{Miller2015} {{Miller}, A.~A. and {Bloom}, J.~S. and {Richards}, J.~W. and 
	{Lee}, Y.~S. and {Starr}, D.~L. and {Butler}, N.~R. and {Tokarz}, S. and 
	{Smith}, N. and {Eisner}, J.~A.}, 2014, ApJ, 798, 122.
\bibitem{SNEx} {{Howell}, A. and {Valenti}, S. and {McCully}, C. and {Arcavi}, I. and {Hosseinzadeh}, G.}, {https://supernova.exchange/public/}. 
\bibitem{RoboNet} {{Street}, R.A. and {Tsapras}, Y. and {Hundertmark}, M. and {Bachelet}, E.}, {https://robonet.lco.global/db}. 
\bibitem{NEOExchange} {{Lister}, T.A. and {Greenstreet}, S. and {Chatelain}, J.}, {https://lco.global/neoexchange/}.
\bibitem{Tsapras2009RBN} {{Tsapras}, Y. and {Street}, R. and {Horne}, K. and {Snodgrass}, C. and 
	{Dominik}, M. and {Allan}, A. and {Steele}, I. and {Bramich}, D.~M. and 
	{Saunders}, E.~S. and {Rattenbury}, N. and {Mottram}, C. and 
	{Fraser}, S. and {Clay}, N. and {Burgdorf}, M. and {Bode}, M. and 
	{Lister}, T.~A. and {Hawkins}, E. and {Beaulieu}, J.~P. and 
	{Fouqu{\'e}}, P. and {Albrow}, M. and {Menzies}, J. and {Cassan}, A. and 
	{Dominis-Prester}, D.},2009, AN, 330, 4.
\bibitem{Hundertmark2018} {{Hundertmark}, M. and {Street}, R.~A. and {Tsapras}, Y. and 
	{Bachelet}, E. and {Dominik}, M. and {Horne}, K. and {Bozza}, V. and 
	{Bramich}, D.~M. and {Cassan}, A. and {D'Ago}, G. and {Figuera Jaimes}, R. and 
	{Kains}, N. and {Ranc}, C. and {Schmidt}, R.~W. and {Snodgrass}, C. and 
	{Wambsganss}, J. and {Steele}, I.~A. and {Mao}, S. and {Ment}, K. and 
	{Menzies}, J. and {Li}, Z. and {Cross}, S. and {Maoz}, D. and 
	{Shvartzvald}, Y.}, 2018, A\&A, 609, A55.
\bibitem{Bramich2008} {{Bramich}, D.~M.}, 2008, MNRAS, 386, L77.
\bibitem{Bramich2013} {{Bramich}, D.~M. and {Horne}, K. and {Albrow}, M.~D. and {Tsapras}, Y. and 
	{Snodgrass}, C. and {Street}, R.~A. and {Hundertmark}, M. and 
	{Kains}, N. and {Arellano Ferro}, A. and {Figuera}, J.~R. and 
	{Giridhar}, S.}, 2013, MNRAS, 428, 2275.
\bibitem{DanDIA} {{Bramich}, D.}, 2017,  {http://adsabs.harvard.edu/abs/2017ascl.soft09005B}. 

\end{thebibliography}
\end{document}